\documentclass[physrev]{revtex4-2}

\usepackage{graphicx}
\usepackage{dcolumn}
\usepackage{bm}
\usepackage{centernot}
\usepackage{amsmath}

\begin{document}

\title{\textbf{Improper currents in theories with local invariance} 
}

\author{Nuno Barros e S\'{a}}
\affiliation{\small FCT, Universidade dos A\c{c}ores, 9500-321 Ponta Delgada, Portugal}
\affiliation{\small Instituto de Astrof\'{i}sica e Ci\^{e}ncias do Espa\c{c}o, Faculdade de Ci\^{e}ncias da Universidade de Lisboa, Edif\'{i}cio C8, Campo Grande, P-1749-016 Lisbon, Portugal.}

\date{\today}

\begin{abstract}
We present a proof that the currents arising from Noether's first theorem in a physical theory with local invariance can always be decomposed into two terms, one of them vanishing on-shell, and the other having an off-shell vanishing divergence, or that they are improper, using the original terminology of Noether. We also prove that, when there is a current which is covariantly conserved, it differs from the canonical current by an improper current. Both proofs are performed in the most general case, that is, for arbitrary maximal order of the derivatives of the dynamical fields of the theory in the Lagrangian, and for arbitrary maximal order of the derivatives of the parameters of the symmetry transformations present in the infinitesimal transformations of the fields and spacetime coordinates. Both proofs are made using only elementary calculus, making them accessible to a large number of physicists. 
\end{abstract}


\maketitle

\section{Introduction}

Noether's first theorem states that, when a physical theory is invariant (in the sense that its equations of motion remain unaltered) under a $A$-dimensional group of transformations with parameters $\varepsilon^a$, there are $A$ local functions $j_a^\mu$ of the dynamical fields of the theory whose divergences are linear combinations of the equations of motion, and hence vanish on-shell, $\partial_\mu j_a^\mu\sim 0$.

Noether's second theorem states that when these transformations are local, that is, when the parameters of the transformation are allowed to vary from point to point in spacetime, $\varepsilon^a(x)$, and the theory is still invariant this enlarged (infinite-dimensional) symmetry group, then there are $A$ relations between the equations of motion of the theory.

However, the infinite-dimensional local group of transformations certainly contains finite-dimensional subgroups obtained by specializing the $g$ functions $\varepsilon^a(x)$ that parameterize the group (such as taking all the parameters to be constants). Therefore, in the case of local transformations, the first theorem can still be applied to such specializations to produce conserved currents $J^\mu(\varepsilon^a)$. In fact, one can in this way generate infinite conserved currents, as there are infinite ways to choose functions $\varepsilon^a(x)$.

In her famous article \cite{noe}, Noether showed that conserved currents generated in this way from local invariance groups can always be decomposed into two parcels,
\begin{equation}
J^\mu(\varepsilon^a)=\Psi^\mu(\varepsilon^a)+R^\mu(\varepsilon^a)\label{umo}\text{ ,}
\end{equation}
one of them vanishing on-shell, $\Psi^\mu(\varepsilon^a)\sim 0$, and the other having a vanishing divergence off-shell, $\partial_\mu R^\mu(\varepsilon^a)=0$. This means that these currents are numerically equal (because they differ by terms that are null when the equations of motion are satisfied) to trivial currents (in the sense that their divergence is mathematically null and hence produces no physical relevant information). Noether called such currents ``improper'', reserving the name ``proper'' for currents for which such decomposition is not possible, as happens in the case of theories with only global invariance. Modern literature often uses the word ``trivial'' \cite{bbh,olv} to mean currents with vanishing off-shell divergence, and sometimes the word ``strong'' is used in that sense \cite{gol}. We shall stick to Noether's original word ``improper'' to mean a current that can be decomposed in the manner of Eq. \eqref{umo}.

Noether proved this assertion using general arguments based on the converse theorems (Section 6 of \cite{noe}). Since then, other proofs have been put forward, some of them not fully general. Most of them involve more sophisticated mathematics, like Ref.\cite{bbh} (Section 6), where the reader must first be introduced to the concepts of Noether operator, cohomology, the Koszul-Tate differential, etc., or Ref. \cite{olv} where the reader is required to go through the concepts of generalized vector fields, adjoint of a differential operator, characteristics and the Frechet derivative. We are not aware of the existence of an explicit derivation of this result which is completely general and simple in its mathematical formulation.

It is therefore the purpose of this article to provide a general proof, that is, valid for arbitrary maximal order $n$ of the derivatives of the fields appearing in the Lagrangian of a theory and for arbitrary maximal order $m$ of the derivatives of the parameters of the invariance group appearing in the infinitesimal transformation law of the dynamical fields, which is simple enough to be understood by any physicist with no mathematical effort. Indeed, the demonstration that follows can be understood simply with the baggage of an undergraduate course on calculus (in fact, very much in the same spirit of the original article by Noether).

The results of this demonstration can be further applied to the case of physical theories which arise from the extension of a global invariance to local status by the addition of extra dynamical fields (gauge fields) such that to the properly conserved currents (canonical) of the original globally invariant theory correspond laws of covariant conservation in the locally invariant theory. In this situation we demonstrate, again in the most general situation, that the canonical current $j_a^\mu$ always equals the covariantly conserved current ${\cal J}_a^\mu$,
\begin{equation}
    j_a^{\mu}={\cal J}_a^{\mu}+\Psi_a^{\mu}-\partial_\chi{\cal S}_a^{\mu\chi}\text{ ,}
\end{equation}
up to a term that vanishes on-shell, $\Psi_a^{\mu}\sim 0$, plus the derivative of a superpotential such that its divergence vanishes off-shell, $\partial_\mu\partial_\chi{\cal S}_a^{\mu\chi}$. Or, putting it another way, that the canonical current and the covariantly conserved current differ at most by an improper current. 

\section{Noether's theorems}

Given the integral $S$ in $D$ dimensions of a functional $L$ of the fields $\phi _{I}$ (with $I=1,\ldots,N$), which is local, meaning that it may depend on these $N$ fields and their derivatives up to a finite order $n$,
\begin{equation}
S=\int L\left( \phi _{I}\right) d^Dx \text{ ,}
\end{equation}%
an infinitesimal variation of the fields $\delta \phi _{I}$ and of the space-time coordinates $\delta x^\mu$ produces the following variation in the integral \cite{noe}
\begin{equation}
\delta S=\int\left(\sum_{I=1}^N\frac{\delta L}{\delta \phi _{I}}\delta \phi _{I}-\partial_\mu B^\mu\right)d^Dx \text{ ,}
\label{bum}
\end{equation}
where
\begin{equation}
    B^{\mu_1}=\sum_{I=1}^N\sum_{j=1}^n\sum_{i=1}^j\left( -1\right) ^i{j\choose i}\partial_{\mu_i}\ldots\partial_{\mu_2}\left[\frac{\partial L}{\partial\left(\partial_{\mu_j}\ldots\partial_{\mu_1}\phi_I\right)}\partial_{\mu_j}\ldots\partial_{\mu_{i+1}}\delta\phi_I\right]-L\,\delta x^{\mu_1}\text{ .}\label{cur}
\end{equation}
Here, and in the remaining of the article, it is understood that indices must be numbered from right to left, that is, $\partial_{\mu_3}\ldots\partial_{\mu_1}\phi=\partial_{\mu_3}\partial_{\mu_2}\partial_{\mu_1}\phi$, $\partial_{\mu_1}\ldots\partial_{\mu_1}\phi=\partial_{\mu_1}\phi$, but $\partial_{\mu_1}\ldots\partial_{\mu_2}\phi=\phi$\text{ .}
We also use the notation for the functional derivative, somewhat liberally, to represent the Lagrange expressions, that is
\begin{equation}
\frac{\delta L}{\delta \phi _{I}}=\sum_{i=0}^{n}\left( -1\right) ^{i}\partial _{\mu _{i}}\cdots \partial
_{\mu _{1}}\frac{\partial L}{\partial \left( \partial _{\mu _{i}}\cdots
\partial _{\mu _{1}}\phi _{I}\right)} \text{ .}
\end{equation}

When the integral $S$ represents the action for a field theory, the equations of motion are found by requiring that they provide an extremum for the action, when the fields are fixed at the boundary of the domain of integration, that is, that $\delta S=0$ when $\delta \phi _{I}=0$ at the boundary and $\delta x^\mu=0$. This implies that the Lagrange expressions vanish for the solutions of the problem, a condition often called ``on-shell'', and that we shall represent by the symbol $\sim$,
\begin{equation}
\frac{\delta L}{\delta \phi _{I}}\sim 0\text{ .}
\end{equation}

Now, let us consider a group of local transformations parameterized by a finite number of functions $\varepsilon _{a}$ ($a=1,\ldots,A$) such that an infinitesimal transformation of the fields depends on $\varepsilon _{a}$ and their derivatives up to some finite order $m$,
\begin{equation}
\delta \phi _{I} =\sum_{i=0}^{m}f_{Ia}^{\mu _{i}\ldots \mu _{1}}\partial _{\mu _{i}}\cdots
\partial _{\mu _{1}}\varepsilon ^{a}\text{ ,}\label{transf}
\end{equation}
where $f_{Ia}^{\mu _{i}\ldots \mu _{1}}$ are some functions, possibly
depending on the fields $\phi _{I}$ and their derivatives, and such that the variation of the action resulting from such transformation amounts at most to a boundary term, dependent on the fields $\phi_I$ and on the parameters $\varepsilon^a$,
\begin{equation}
\delta S=\int \partial_\mu K^\mu\, d^Dx\text{ .}
\end{equation}
These transformations are called symmetries of the action because they leave the equations of motion invariant.

Then, Eq. \eqref{bum} becomes
\begin{equation}
\int\left(\sum_{I=1}^N\frac{\delta L}{\delta \phi _{I}}\delta \phi _{I}-\partial_\mu J^\mu\right)d^Dx =0\text{ ,}\label{bumbum}
\end{equation}
with $\delta \phi _{I}$ given by Eq. \eqref{transf} and
\begin{equation}
J^\mu=B^\mu+K^\mu\text{ .}\label{bumba}
\end{equation}
Because this result is valid for any domain of integration, we conclude that
\begin{equation}
\partial_\mu J^\mu=\sum_{I=1}^N\frac{\delta L}{\delta \phi _{I}}\delta \phi _{I}\sim 0\text{ .}
\end{equation}
This is the equation that provides the conserved currents of Noether's first theorem.

Since $\varepsilon ^{a}$ are arbitrary functions, one can choose them such that their derivatives up to a sufficiently high finite order vanish at the boundary of the domain of integration, in which case Eq. \eqref{bumbum} becomes, after further integration by parts,
\begin{equation}
\int\sum_{I=1}^N\sum_{i=0}^{m}\left( -1\right) ^{i}\partial _{\mu
_{i}}\cdots \partial _{\mu _{1}}\left( \frac{\delta L}{\delta \phi _{I}}%
f_{Ia}^{\mu _{i}\ldots \mu _{1}}\right)\varepsilon^{a} d^Dx =0 \text{ ,}
\end{equation}
or,
\begin{equation}
\sum_{I=1}^N\sum_{i=0}^{m}\left( -1\right) ^{i}\partial _{\mu _{i}}\cdots \partial _{\mu
_{1}}\left( \frac{\delta L}{\delta \phi _{I}}f_{Ia}^{\mu _{i}\ldots \mu
_{1}}\right) =0\text{ .}  \label{bquatro}
\end{equation}%
This is the content of Noether's second theorem: there are $A$ relations among the equations of motion. They reflect the fact that the evolution of the system is not well-defined, as one should expect since all field configurations related by a symmetry transformation represent the same physical state.

\section{Improperness of currents}

Since $J^\mu$ is of order $n-1$ in the derivatives of the variations of the fields $\delta\phi_I$, which are themselves of order $m$, the current $J^\mu$ is of order $m+n-1$ in the derivatives of the parameters $\varepsilon^{a}$ of the transformations,
\begin{equation}
J^{\mu_0}=\sum_{i=0}^{m+n-1}j_a^{\mu_i\ldots\mu_0}\partial_{\mu_i}\ldots\partial_{\mu_1}\varepsilon^{a}\text{ .}\label{dof}
\end{equation}
Then, we can expand the integrand of Eq. \eqref{bum} in orders of the derivatives of the parameters of the transformation, with the result
\begin{eqnarray}
    &&\sum_{I=1}^N\frac{\delta L}{\delta \phi _{I}}\delta \phi _{I}-\partial_\mu J^\mu=\left(\sum_{I=1}^N\frac{\delta L}{\delta \phi _{I}}f_{Ia}-\partial_{\mu_0}j_a^{\mu_0}\right)\varepsilon^{a}
    +\nonumber\\
    &&+\sum_{i=1}^{m}\left(\sum_{I=1}^N\frac{\delta L}{\delta \phi _{I}}f_{Ia}^{\mu _{i}\ldots \mu _{1}}-\partial_{\mu_0}j_a^{\mu_i\ldots\mu_0}-j_a^{\mu_i\ldots\mu_1}\right)\partial _{\mu _{i}}\cdots\partial _{\mu _{1}}\varepsilon^{a}-\nonumber\\
    &&-\sum_{i=m+1}^{m+n-1}\left(\partial_{\mu_0}j_a^{\mu_i\ldots\mu_0}+j_a^{\mu_i\ldots\mu_1}\right)\partial_{\mu_i}\ldots\partial_{\mu_1}\varepsilon _{a}-j_a^{\mu_{m+n}\ldots\mu_1}\partial_{\mu_{m+n}}\ldots\partial_{\mu_1}\varepsilon^{a}\text{ .}
\end{eqnarray}
If the action is to be invariant under all transformations, the coefficients of this expansion must all vanish, hence
\begin{eqnarray}
    0&=&\sum_{I=1}^N\frac{\delta L}{\delta \phi _{I}}f_{Ia}-\partial_{\nu}j_a^{\nu}\label{a1}\\
    j_a^{\left(\mu_i\ldots\mu_0\right)}&=&\sum_{I=1}^N\frac{\delta L}{\delta \phi _{I}}f_{Ia}^{\mu _{i}\ldots \mu _{0}}-\partial_{\nu}j_a^{\mu_i\ldots\mu_0\nu}=0\quad {\rm for}\quad i=0,\ldots,m-1\label{a2}\\
    j_a^{\left(\mu_i\ldots\mu_0\right)}&=&-\partial_{\nu}j_a^{\mu_i\ldots\mu_0\nu}\quad {\rm for}\quad i=m,\ldots,m+n-2\\
    j_a^{\left(\mu_{m+n-1}\ldots\mu_0\right)}\label{a4}&=&0\text{ .}
\end{eqnarray}
The brackets involving the indices on the left-hand sides of these equations mean total symmetrization.

The functions $j_a^{\mu_i\ldots\mu_0}$ are, by their definition Eq. \eqref{dof}, symmetric in all their indices, except for the last one, We shall split each of them in the totally symmetric part and the remaining part,
\begin{equation}
j_a^{\mu_i\ldots\mu_0}=j_a^{\left(\mu_i\ldots\mu_0\right)}+R_a^{\mu_i\ldots\mu_0}\text{ ,}
\end{equation}
with
\begin{equation}
R_a^{\left(\mu_i\ldots\mu_0\right)}=0\text{ .}
\end{equation}

Then, Eqs. \eqref{a2}-\eqref{a4} become
\begin{eqnarray}
    j_a^{\left(\mu_i\ldots\mu_0\right)}&=&\sum_{I=1}^N\frac{\delta L}{\delta \phi _{I}}f_{Ia}^{\mu _{i}\ldots \mu _{0}}-\partial_{\nu}j_a^{\left(\mu_i\ldots\mu_0\nu\right)}-\partial_{\nu}R_a^{\mu_i\ldots\mu_0\nu}\quad {\rm for}\quad i=0,\ldots,m-1\\
    j_a^{\left(\mu_i\ldots\mu_0\right)}&=&-\partial_{\nu}j_a^{\left(\mu_i\ldots\mu_0\nu\right)}-\partial_{\nu}R_a^{\mu_i\ldots\mu_0\nu}\quad {\rm for}\quad i=m,\ldots,m+n-2\\
    j_a^{\left(\mu_{m+n-1}\ldots\mu_0\right)}&=&0\text{ .}
\end{eqnarray}
This system can be solved for the symmetric parts $j_a^{\left(\mu_i...\mu_0\right)}$, producing
\begin{eqnarray}
    j_a^{\left(\mu_i\ldots\mu_0\right)}&=&\sum_{j=1}^{m+n-1-i}\left(-1\right)^j\partial_{\nu_j}\ldots\partial_{\nu_1}R_a^{\mu_i\ldots\mu_0\nu_j\ldots\nu_1}-\nonumber\\
    &&-\sum_{j=1}^{m-i}\left(-1\right)^j\partial_{\nu_j}\ldots\partial_{\nu_2}\left(\sum_{I=1}^N\frac{\delta L}{\delta \phi _{I}}f_{Ia}^{\mu _{i}\ldots \mu _{0}\nu_j\ldots\nu_2}\right)\quad {\rm for}\quad i=0,\ldots,m-1\\
    j_a^{\left(\mu_i\ldots\mu_0\right)}&=&\sum_{j=1}^{m+n-1-i}\left(-1\right)^j\partial_{\nu_j}\ldots\partial_{\nu_1}R_a^{\mu_i\ldots\mu_0\nu_j\ldots\nu_1}\quad {\rm for}\quad i=m,\ldots,m+n-1\text{ .}
\end{eqnarray}

Finally, adding the remaining parts $R_a^{\mu_i...\mu_0}$ to the previous solution, one gets
\begin{eqnarray}
    j_a^{\mu_i\ldots\mu_0}&=&\sum_{j=0}^{m+n-1-i}\left(-1\right)^j\partial_{\nu_j}\ldots\partial_{\nu_1}R_a^{\mu_i\ldots\mu_0\nu_j\ldots\nu_1}-\nonumber\\
    &&-\sum_{j=1}^{m-i}\left(-1\right)^j\partial_{\nu_j}\ldots\partial_{\nu_2}\left(\sum_{I=1}^N\frac{\delta L}{\delta \phi _{I}}f_{Ia}^{\mu _{i}\ldots \mu _{0}\nu_j\ldots\nu_2}\right)\quad {\rm for}\quad i=0,\ldots,m-1\label{b1}\\
    j_a^{\mu_i\ldots\mu_0}&=&\sum_{j=0}^{m+n-1-i}\left(-1\right)^j\partial_{\nu_j}\ldots\partial_{\nu_1}R_a^{\mu_i\ldots\mu_0\nu_j\ldots\nu_1}\quad {\rm for}\quad i=m,\ldots,m+n-1\text{ .}\label{b2}
\end{eqnarray}

One can use Eq. \eqref{a1} to check the consistency of the solution Eqs. \eqref{b1}-\eqref{b2}. We have
\begin{equation}
j_a^{\mu_0}=\sum_{j=0}^{m+n-1}\left(-1\right)^j\partial_{\nu_j}\ldots\partial_{\nu_1}R_a^{\mu_0\nu_j\ldots\nu_1}-\sum_{j=1}^{m}\left(-1\right)^j\partial_{\nu_j}\ldots\partial_{\nu_2}\left(\sum_{I=1}^N\frac{\delta L}{\delta \phi _{I}}f_{Ia}^{\mu _{0}\nu_j\ldots\nu_2}\right)\text{ .}\label{b3}
\end{equation}
Therefore
\begin{equation}
\partial_{\mu_0}j_a^{\mu_0}=-\sum_{j=1}^{m}\left(-1\right)^j\partial_{\nu_j}\ldots\partial_{\nu_1}\left(\sum_{I=1}^N\frac{\delta L}{\delta \phi _{I}}f_{Ia}^{\nu_j\ldots\nu_1}\right)\text{ .}
\end{equation}
Plugging Eq. \eqref{a1} in the left-hand side of this equation, one simply recovers Noether's second theorem, Eq. \eqref{bquatro}.

Using the solution Eqs. \eqref{b1}-\eqref{b2}, we can finally rewrite the Noether current Eq. \eqref{dof} in the form
\begin{equation}
J^{\mu_0}=\Psi^{\mu_0}+R^{\mu_0}\text{ ,}
\end{equation}
with
\begin{eqnarray}
\Psi^{\mu_0}&=&\sum_{i=0}^{m+n-1}\sum_{j=0}^{m-i-1}\left(-1\right)^j\partial_{\nu_j}\ldots\partial_{\nu_1}\left(\sum_{I=1}^N\frac{\delta L}{\delta \phi _{I}}f_{Ia}^{\mu _{i}\ldots\mu _{0}\nu_j\ldots\nu_1}\right)\partial_{\mu_i}\ldots\partial_{\mu_1}\varepsilon^{a}\\
R^{\mu_0}&=&\sum_{i=0}^{m+n-1}\sum_{j=0}^{m+n-1-i}\left(-1\right)^j\partial_{\nu_j}\ldots\partial_{\nu_1}R_a^{\mu_i\ldots\mu_0\nu_j\ldots\nu_1}\partial_{\mu_i}\ldots\partial_{\mu_1}\varepsilon^{a}\text{ .}
\end{eqnarray}
The quantity $\Psi^{\mu_0}$ is a combination of the equations of motion and their derivatives, hence it vanishes on-shell, $\Psi^{\mu_0}\sim 0$.

And we can verify that the divergence of the remaining part of the current, $R^{\mu_0}$, vanishes off-shell,
\begin{eqnarray}
\partial_{\mu_0}R^{\mu_0}&=&\sum_{i=0}^{m+n-1}\sum_{j=0}^{m+n-1-i}\left(-1\right)^j\partial_{\nu_j}\ldots\partial_{\nu_0}R_a^{\mu_i\ldots\mu_1\nu_j\ldots\nu_0}\partial_{\mu_i}...\partial_{\mu_1}\varepsilon^{a}+\nonumber\\
&&+\sum_{i=0}^{m+n-1}\sum_{j=0}^{m+n-1-i}\left(-1\right)^j\partial_{\nu_j}\ldots\partial_{\nu_1}R_a^{\mu_i\ldots\mu_0\nu_j\ldots\nu_1}\partial_{\mu_i}...\partial_{\mu_0}\varepsilon _{a}=\nonumber\\
&=&\sum_{i=1}^{m+n-1}\sum_{j=0}^{m+n-1-i}\left(-1\right)^j\partial_{\nu_j}\ldots\partial_{\nu_0}R_a^{\mu_i\ldots\mu_1\nu_j\ldots\nu_0}\partial_{\mu_i}...\partial_{\mu_1}\varepsilon^{a}+\nonumber\\
&&+\sum_{i=0}^{m+n-2}\sum_{j=1}^{m+n-1-i}\left(-1\right)^j\partial_{\nu_j}\ldots\partial_{\nu_1}R_a^{\mu_i\ldots\mu_0\nu_j\ldots\nu_1}\partial_{\mu_i}...\partial_{\mu_0}\varepsilon^{a}=0
\end{eqnarray}

\section{Covariantly conserved currents}

Let us now specialize to the case of non-abelian gauge theories when the first $M$ fields $\phi_I$ are the gauge fields $A_\mu^a$, and the Lagrangian splits in the form
\begin{equation}
    L=L_g(A_\mu^a)+L_m(A_\mu^a,\phi_{M+1},...\phi_N)\text{ .}
\end{equation}
Then, setting to zero
\begin{equation}
\Psi^I=\frac{\delta L_m}{\delta\phi_I},\quad I=M+1,...,N    
\end{equation}
produces equations of motion, but not
\begin{equation}
{\cal J}_a^\mu=\frac{\delta L_m}{\delta A_\mu^a}\text{ ,}\label{equ3} 
\end{equation}
as the equations of motion for the gauge fields $A_\mu^a$ involve the functional derivative of $L_g$ too.

Next we shall apply the results of the previous section to $S_m=\int L_m d^Dx$ only, assuming that it is invariant on its own under the transformations given by Eq. \eqref{transf}. Equation \eqref{bumbum} becomes
\begin{equation}
\int\left(\frac{\delta L_m}{\delta A^a_\mu}\delta A ^a_\mu+\sum_{I=M}^N\frac{\delta L_m}{\delta \phi _{I}}\delta \phi _{I}-\partial_\mu J^\mu\right)d^Dx=0\text{ ,}\label{plo}
\end{equation}
with $J^{\mu}_m$ given by Eqs. \eqref{cur} and \eqref{bumba}, but with $L$ replaced by $L_m$, and where $K^\mu$ may only arise from the variation of $L_m$. Notice that the sum in $I$ runs from $1$ to $N$, which includes the gauge fields $A_\mu^a$.

The variation of the gauge fields under a gauge transformation parameterized by $\varepsilon^a$ is given by
\begin{equation}
    \delta A_\mu^a=D_\mu\varepsilon^a=\partial_\mu\varepsilon^a+g\,C^{abc}A_\mu^b\varepsilon^c\text{ ,}
\end{equation}
where $g$ is the coupling constant, $C^{abc}$ are the structure constants of the group, and $D_\mu$ represents the covariant derivative. We have then
\begin{equation}
    \frac{\delta L_m}{\delta A^a_\mu}\delta A ^a_\mu=\frac{\delta L_m}{\delta A^a_\mu}D_\mu\varepsilon^a=\partial_\mu\left(\frac{\delta L_m}{\delta A^a_\mu}\varepsilon^a\right)-D_\mu\frac{\delta L_m}{\delta A^a_\mu}\varepsilon^a\text{ .}\label{labu}
\end{equation}

In the case of a diffeomorphism invariant theory where the first $M$ fields are the metric $g_{\mu\nu}$, and
\begin{equation}
-\delta g_{\mu\nu}=\nabla_\mu\varepsilon_\nu+\nabla_\nu\varepsilon_\mu=\left(\Gamma_{\chi\mu\nu}+\Gamma_{\chi\nu\mu}\right)\varepsilon^\chi+g_{\chi\mu}\partial_\nu\varepsilon^\chi+g_{\chi\nu}\partial_\mu\varepsilon^\chi\text{ ,}\label{labu0}
\end{equation}
where $\nabla_\mu$ is the covariant derivative with respect to the Levi-Civita connection $\Gamma_{\mu\nu}{^\chi}$, one has
\begin{equation}
    \frac{\delta L_m}{\delta g_{\mu\nu}}\delta g_{\mu\nu}=-2\frac{\delta L_m}{\delta g_{\mu\nu}}\nabla_\mu\varepsilon_\nu=\partial_\mu\left(-2\frac{\delta L_m}{\delta g_{\mu\nu}}g_{\nu\chi}\varepsilon^a\right)-\nabla_\mu\left(-2\frac{\delta L_m}{\delta g_{\mu\nu}}g_{\nu\chi}\right)\varepsilon^\chi\text{ ,}
\end{equation}
so that $\nabla_\mu$ plays the role of $D_\mu$, and the symmetric tensor
\begin{equation}
    {\cal J}_\chi^\mu=-2\frac{\delta L_m}{\delta g_{\mu\nu}}g_{\nu\chi}
\end{equation}
plays the role of ${\cal J}_a^\mu$. Therefore we shall treat both cases, non-abelian gauge theories and diffeomorphism invariant theories, as one, and simply call ${\cal J}_a^\mu$ the covariantly conserved current.

We have then, from Eqs. \eqref{plo} and \eqref{labu},
\begin{equation}
\int\left[\partial_\mu\left({\cal J}_a^\mu\varepsilon^a\right)-D_\mu{\cal J}_a^\mu\varepsilon^a+\sum_{I=M}^N\Psi^{I}\sum_{i=0}^{m}f_{Ia}^{\mu
_{i}\ldots \mu _{1}}\partial _{\mu _{i}}\cdots \partial _{\mu
_{1}}\varepsilon^{a}-\partial_\mu J^\mu\right]d^Dx=0\text{ .}\label{bil}
\end{equation}
Further integrating by parts and ignoring boundary terms, we get the law of covariant conservation
\begin{equation}
D_\mu{\cal J}_a^\mu=\sum_{I=M}^N\sum_{i=0}^{m}\left(-1\right)^i\partial _{\mu _{i}}\cdots \partial _{\mu
_{1}}\left(\Psi^If_{Ia}^{\mu
_{i}\ldots \mu _{1}}\right)\text{ .}\label{law}
\end{equation}

Plugging Eq. \eqref{law} in Eq. \eqref{bil} we obtain
\begin{eqnarray}
&&\int\left[\partial_\mu\left({\cal J}_a^\mu\varepsilon^a-J^\mu\right)+\sum_{I=M}^N\sum_{i=1}^{m}\Psi^I f_{Ia}^{\mu
_{i}\ldots \mu _{1}}\partial _{\mu _{i}}\cdots \partial _{\mu
_{1}}\varepsilon^{a}-\right.\nonumber\\
&&\left.\sum_{I=M}^N\sum_{i=1}^{m}\left(-1\right)^i\partial _{\mu _{i}}\cdots \partial _{\mu
_{1}}\left(\Psi^I f_{Ia}^{\mu
_{i}\ldots \mu _{1}}\right)\varepsilon^a\right]d^Dx=0\text{ .}
\end{eqnarray}
Setting the coefficients of all orders of derivatives of $\varepsilon^a$ to zero, we get a system very similar to Eqs. \eqref{a1}-\eqref{a4}, the only differences being that the sums over fields go from $M$ to $N$, not from $1$ to $N$, and the replacements 
\begin{eqnarray}
\sum_{I=M}^N\Psi^If_{Ia}\,&\to&\,-\sum_{I=M}^N\sum_{i=1}^{m}\left(-1\right)^i\partial _{\mu _{i}}\cdots \partial _{\mu_{1}}\left(\Psi^I f_{Ia}^{\mu_{i}\ldots \mu _{1}}\right)\quad {\rm in\ Eq.\ \eqref{a1}}\\
j^\mu_a\,&\to&\,j^\mu_a-{\cal J}^\mu_a\text{ .}
\end{eqnarray}
This means that the solutions Eqs. \eqref{b1}-\eqref{b2} are the same, except for the lowest order, Eq. \eqref{b3}, which becomes
\begin{equation}
j_a^{\mu}={\cal J}_a^{\mu}+\Psi_a^{\mu}+S_a^{\mu}\label{e41}\text{ ,}
\end{equation}
with
\begin{eqnarray}
\Psi_a^{\mu}&=&\sum_{j=0}^{m-1}\left(-1\right)^j\partial_{\nu_j}\ldots\partial_{\nu_1}\left(\sum_{I=M}^N\Psi^I f_{Ia}^{\mu\nu_j\ldots\nu_1}\right)\\
S_a^{\mu}&=&\sum_{j=1}^{m+n-1}\left(-1\right)^j\partial_{\nu_j}\ldots\partial_{\nu_1}R_a^{\mu\nu_j\ldots\nu_1}\text{ .}
\end{eqnarray}
Therefore, we found out that $j_a^{\mu}$, the term in of lowest in order of the derivatives of the parameters $\varepsilon
^a$ of the current $J^\mu$, equals the covariantly conserved current up to a term that vanishes on-shell ($\Psi_a^{\mu}$) plus a term whose divergence vanishes off-shell ($\partial_{\mu}S_a^{\mu}=0$), because, since the totally symmetric part of $R_a^{\mu\nu_j...\nu_1}$ vanishes, $\partial_{\mu}\partial_{\nu_j}...\partial_{\nu_1}R_a^{\mu\nu_j...\nu_1}=0$ holds.

\section{Canonical currents}

In non-abelian gauge theories, from Eqs. \eqref{cur} and \eqref{bumba} together with Eq. \eqref{labu}, we can read 
\begin{eqnarray}
    j^{\mu_1}_a&=&g\,C^{abc}\sum_{j=1}^n\sum_{i=1}^j\left( -1\right) ^i{j\choose i}\partial_{\mu_i}...\partial_{\mu_2}\left[\frac{\partial L_m}{\partial\left(\partial_{\mu_j}...\partial_{\mu_1}A_\nu^b\right)}\partial_{\mu_j}...\partial_{\mu_{i+1}}A_\mu^c\right]+\nonumber\\
    &&+\sum_{I=M}^N\sum_{j=1}^n\sum_{i=1}^j\left( -1\right) ^i{j\choose i}\partial_{\mu_i}...\partial_{\mu_2}\left[\frac{\partial L_m}{\partial\left(\partial_{\mu_j}...\partial_{\mu_1}\phi_I\right)}\partial_{\mu_j}...\partial_{\mu_{i+1}}f_{Ia}\right]+\frac{\partial K^{\mu_1}}{\partial \varepsilon^a}\text{ .}\label{labo1}
\end{eqnarray}

In diffeomorphism invariant theories, from Eq. \eqref{cur} and \eqref{bumba} together with Eq. \eqref{labu0} and $\delta x^\mu=\varepsilon^\mu$, we get
\begin{eqnarray}
    j^{\mu_1}_{\nu_1}&=&\sum_{j=1}^n\sum_{i=1}^j\left( -1\right) ^i{j\choose i}\partial_{\mu_i}...\partial_{\mu_2}\left[\frac{\partial L_m}{\partial\left(\partial_{\mu_j}...\partial_{\mu_1}g_{\nu_2\nu_3}\right)}\partial_{\mu_j}...\partial_{\mu_{i+1}}\left(\Gamma_{\nu_1\nu_2\nu_3}+\Gamma_{\nu_1\nu_2\nu_3}\right)\right]+\nonumber\\
    &&+\sum_{I=M}^N\sum_{j=1}^n\sum_{i=1}^j\left( -1\right) ^i{j\choose i}\partial_{\mu_i}...\partial_{\mu_2}\left[\frac{\partial L_m}{\partial\left(\partial_{\mu_j}...\partial_{\mu_1}\phi_I\right)}\partial_{\mu_j}...\partial_{\mu_{i+1}}f_{I\nu_1}\right]-L\delta_{\nu_1}^{\mu_1}+\frac{\partial K^{\mu_1}}{\partial \xi^{\nu_1}}\text{ .}\label{labo2}
\end{eqnarray}

Setting $A_\mu^a=0$ or $g_{\mu\nu}=\eta_{\mu\nu}$ in the matter part of the action $S_m$, produces an integral which is invariant under global transformations only. In such cases there are currents which are properly conserved, and they can be obtained from Eqs. \eqref{labo1} and \eqref{labo2} by setting respectively $A_\mu^a=0$ or $g_{\mu\nu}=\eta_{\mu\nu}$. Notice that the first terms in both equations vanish when $A_\mu^a=0$ or $g_{\mu\nu}=\eta_{\mu\nu}$. They are called canonical currents.

In the context of General Relativity, the current obtained by setting $g_{\mu\nu}=\eta_{\mu\nu}$ in the covariantly conserved current ${\cal J}^{\mu}_\nu$ is called the metric energy-momentum tensor. In order to embrace both situations, of gauge theories and of diffeomorphism invariant theories, we shall call it the ``gauge current'', meaning that $A_\mu^a=0$ or $g_{\mu\nu}=\eta_{\mu\nu}$ was set in $g_{\mu\nu}=\eta_{\mu\nu}$, according to the case.

Setting $A_\mu^a=0$ in \eqref{e41}, we see that the gauge current equals the canonical current up to a term that vanishes on-shell plus a term whose divergence vanishes off-shell. Put it another way, they differ by an improper current. Another, more common, way to write Eq. \eqref{e41} is
\begin{equation}
j_a^{\mu}={\cal J}_a^{\mu}+\Psi_a^{\mu}-\partial_\chi{\cal S}_a^{\mu\chi}\text{ ,}
\end{equation}
with
\begin{eqnarray}
\Psi_a^{\mu}&=&\sum_{j=0}^{m-1}\left(-1\right)^j\partial_{\nu_j}...\partial_{\nu_1}\left(\sum_{I=M}^N\Psi^I f_{Ia}^{\mu\nu_j...\nu_1}\right)\\
{\cal S}_a^{\mu\chi}&=&\sum_{j=0}^{m+n-2}\left(-1\right)^j\partial_{\nu_j}...\partial_{\nu_1}R_a^{\mu\left(\nu_j...\nu_1\chi\right)}\text{ .}
\end{eqnarray}
Notice that, when setting $A_\mu^a=0$ in the Lagrange expressions $\Psi^I$ for the theory with local invariance, one gets the Lagrange expressions for the theory with global invariance. The quantity ${\cal S}_a^{\mu\chi}$ is called the superpotential.

In Yang-Mills theory $n=1$ and $m=0$ for the matter terms. Therefore $\Psi_a^{\mu}=0$ and ${\cal S}_a^{\mu\chi}=0$, and the gauge and canonical currents are exactly equal, $j_a^{\mu}={\cal J}_a^{\mu}$.

In General Relativity $n=1$ and $m=1$ for the matter terms. Therefore
\begin{eqnarray}
\Psi_a^{\mu}&=&\sum_{I=M}^N\Psi^I f_{Ia}^{\mu}\\
{\cal S}_a^{\mu\chi}&=&R_a^{\mu\chi}\text{ .}
\end{eqnarray}
This is a well-known result, in particular for the energy-momentum tensor of the electromagnetic field, in which case the reader can easily verify that, for the Lagrangian
\begin{equation}
L_m=-\frac{1}{4}F^2=-\frac{1}{4}F_{\mu\nu}F^{\mu\nu}\text{ ,}
\end{equation}
one has
\begin{eqnarray}
    {\cal J}_\nu^{\mu}&=&\frac{1}{4}F^2\delta_\nu^\mu-F^{\mu\chi}F_{\nu\chi}\\
    j_\nu^{\mu}&=&\frac{1}{4}F^2\delta_\nu^\mu-F^{\mu\chi}\partial_\nu A_\chi\text{ ,}
\end{eqnarray}
and
\begin{equation}
    j_\nu^{\mu}={\cal J}_\nu^{\mu}-A_{\nu}\frac{\delta L}{\delta A_{\mu} }-\partial_\chi\left( F^{\mu\chi}A_\nu\right)\text{ .}
\end{equation}

In the case of General Relativity the superpotential is antisymmetric, because the totally symmetric part of $R_a^{\mu\chi}$ vanishes. But it should be noted that for higher values of $m$ and $n$ that does not need to be the case. While
\begin{equation}
{\cal S}_a^{\mu\chi}+{\cal S}_a^{\chi\mu}=0\implies \partial_\mu\partial_\chi{\cal S}_a^{\mu\chi}=0\text{ ,}
\end{equation}
the converse is not true,
\begin{equation}
\partial_\mu\partial_\chi{\cal S}_a^{\mu\chi}=0\centernot\implies{\cal S}_a^{\mu\chi}+{\cal S}_a^{\chi\mu}=0\text{ .}
\end{equation}
It is worth giving a practical example, because the antisymmetry of the superpotential is frequently stated in the context of General Relativity, without pointing out that it is not a general result. Let
\begin{equation}
    {\cal S}_a^{\mu\chi}=\partial_\nu\left( X^\mu X^\nu Y^\chi_a+X^\chi X^\nu Y^\mu_a- 2X^\mu X^\chi Y^\nu_a\right)\text{ .}
\end{equation}
This tensor is clearly not antisymmetric (in fact, it is symmetric), nevertheless $\partial_\mu\partial_\chi{\cal S}_a^{\mu\chi}=0$ holds.

The term proportional to the equations of motion, $\Psi_\nu^{\mu}$, can be computed exactly for General Relativity assuming that the fields $\phi_I$ are tensor fields (spinor fields require a different treatment). Representing these fields explicitly with their spacetime indices $\phi_{\alpha_1 ...\alpha_{r_I}}^{\beta_1 ...\beta_{s_I}}$, their variations under diffeomorphisms are given by the symmetric of their Lie derivatives,
\begin{equation}
    \delta \phi_{\alpha_1\ldots\alpha_{r_I}}^{\beta_1\ldots\beta_{s_I}}=-\xi^\mu\partial_\mu \phi_{\alpha_1\ldots\alpha_{r_I}}^{\beta_1\ldots\beta_{s_I}}-\sum_{i=1}^{r_I}\phi_{\alpha_1\ldots(\alpha_i\to\mu)\ldots\alpha_{r_I}}^{\beta_1\ldots\beta_{s_I}}\partial_{\alpha_i}\xi^\mu+\sum_{i=1}^{s_I}\phi_{\alpha_1\ldots\alpha_{r_I}}^{\beta_1\ldots(\beta_i\to\mu)\ldots\beta_{s_I}}\partial_\mu\xi^{\beta_i} \text{ .}
\end{equation}
We have then
\begin{equation}
    \Psi_\nu^{\mu}=\sum_{I=M+1}^N\left(\sum_{i=1}^{s_I}\phi_{\alpha_1\ldots\alpha_{r_I}}^{\beta_1\ldots(\beta_i\to\mu)\ldots\beta_{s_I}}\frac{\delta L_m}{\delta\phi_{\alpha_1\ldots\alpha_{r_I}}^{\beta_1\ldots(\beta_i\to\nu)\ldots\beta_{s_I}} }-\sum_{i=1}^{r_I}\phi_{\alpha_1\ldots(\alpha_i\to\nu)\ldots\alpha_{r_I}}^{\beta_1\ldots\beta_{s_I}}\frac{\delta L_m}{\delta \phi_{\alpha_1\ldots(\alpha_i\to\mu)\ldots\alpha_{r_I}}^{\beta_1\ldots\beta_{s_I}} }\right)\text{ ,}\label{sup}
\end{equation}
which is non-zero for all fields other than scalars.

For the construction of the superpotential in General Relativity, and in theories with higher order $n$ of the derivatives of the fields, please see Ref \cite{ilin}.

\end{document}